\documentclass[12pt,journal,draftclsnofoot,letterpaper,onecolumn]{IEEEtran}
\usepackage{url}             % example: \url{my_url_here}.
\usepackage[cmex10]{amsmath}
\usepackage{amsfonts}
\usepackage{amssymb}
\usepackage{cite}
\usepackage[caption=false, font=footnotesize]{subfig}

\ifCLASSINFOpdf
  \usepackage[pdftex]{graphicx}
  \DeclareGraphicsExtensions{.pdf,.jpeg,.jpg,.png}
\else
  \usepackage[dvips]{graphicx}
  \DeclareGraphicsExtensions{.eps}
\fi

\begin{document}

%% can use linebreaks \\ within to get better formatting as desired
\title{Compressed Relaying for Two-Way Relay Networks with Correlated Sources}

\author{Qiang~Huo,
        Kun~Yang,
        Lingyang~Song,
        Yonghui~Li,
        and~Bingli~Jiao%
\thanks{Q.~Huo is with the Physical Layer \& RRM IC Algorithm Department,
Shanghai Research Center,
Huawei Technologies Co., Ltd., Shanghai 201206,
China (e-mail: qianghuoee@gmail.com).}%
\thanks{K.~Yang, L.~Song, and B.~Jiao are with the School of Electronics Engineering and Computer Science, Peking University, Beijing 100871, China (e-mail: \{kun.yang, lingyang.song, jiaobl\}@pku.edu.cn).}%
\thanks{Y.~Li is with the School of Electrical and Information Engineering, the University of Sydney, Sydney, NSW 2006, Australia (e-mail: yonghui.li@sydney.edu.au).}%
}

%% title and authors
%% ==================================== %%

%% ==================================== %%
%% paper headers and etc.

%% The paper headers
%% ----------------------
\markboth{IEEE WIRELESS COMMUNICATIONS LETTERS, ACCEPTED FOR PUBLICATION}%
{Huo \MakeLowercase{\textit{et al.}}: Compressed Relaying for Two-Way Relay Networks with Correlated Sources}
%% ----------------------
%IEEE TRANSACTIONS ON COMMUNICATIONS,~Vol.~6, No.~1, January~2012
%IEEE TRANSACTIONS ON COMMUNICATIONS, ACCEPTED FOR PUBLICATION

%% put a publisher's ID mark on the page
%% Remember, if you use this you must call \IEEEpubidadjcol in the second
%% column for its text to clear the IEEEpubid mark.

%% use for special paper notices

%% make the title area
%% ----------------------
\maketitle
%% ----------------------

%% paper headers and etc.
%% ==================================== %%

%% ==================================== %%
%% abstract section
\begin{abstract}
%\boldmath
%The abstract goes here.
In this letter, a compressed relaying scheme via  Huffman and physical-layer network coding (HPNC) is proposed for two-way relay networks with correlated sources (TWRN-CS).
In the HPNC scheme, both sources first transmit the correlated raw source messages to the relay simultaneously.
The relay performs physical-layer network coding (PNC) on the received symbols, compresses the PNC-coded symbols using Huffman coding, and broadcasts the compressed symbols to both source nodes.
Then, each source decodes the other source's messages by using its own messages as side information.
Compression rate and block error rate (BLER) of the proposed scheme are analyzed.
Simulation results demonstrate that the HPNC scheme can effectively improve the network throughput, and meanwhile, achieve the superior BLER performance compared with the conventional non-compressed relaying scheme in TWRN-CS.
\end{abstract}

%% no keywords in IEEE conference model
\IEEEpeerreviewmaketitle

%% abstract section
%% ==================================== %%

%% ==================================== %%
%% Introduction section
\section{Introduction}\label{sec:introduction}
% \IEEEPARstart{A}{demo} file
% \IEEEPARstart{T}{his} demo file
%% no IEEEPARstart in conference model

\IEEEPARstart{T}{wo-way} relay communication has recently drawn much interest from both academia and industry due to its potential ability to significantly improve the spectral efficiency of wireless communication systems~\cite{Huo2012P4998}.
Particularly, a simple three-node two-way relay network (TWRN) has been widely studied in the literature
and various relaying protocols have been proposed to achieve network throughput improvement~\cite{Wu2004P,Popovski2007P16,Liew2013P4}.

However, most existing relaying protocols for TWRNs have been designed for the networks where the information messages exchanged between two source nodes are independent only.
Recently, some works have reported that measurements between different nodes in the network may exhibit strong correlations in some scenarios, e.g., wireless sensor networks~\cite{Jindal2006P466}.
In such scenarios, an efficient relaying protocol, which can exploit the inherent correlation characteristic, can further improve the system spectral efficiency and performance.
This will be highly desirable when the source nodes want to exchange their correlated information messages with each other via the intermediate relaying nodes.
In~\cite{Timo2013P753}, the authors studied the multi-way relay network with correlated sources and orthogonal uplink channels. The paper mainly focused on the source-channel separation problems from an information theoretical point of view. In this letter, we consider the two-way relay networks with correlated sources (TWRN-CS) and employ   physical-layer network coding~(PNC) and practical Huffman coding ~\cite{Huffman1952P1098} to compress the correlated data in TWRN.

The main contributions and novelties of this letter are two-fold:
1) A compressed relaying scheme via   Huffman and physical-layer network coding (HPNC) is proposed for TWRN-CS;
2) Compression rate is analyzed and a closed-form block error rate (BLER) expression is derived for the HPNC scheme.
It is shown that the HPNC scheme achieves considerable improvements in both network throughput and BLER performance over the conventional
non-compressed relaying scheme in TWRN-CS.

%% Notation
\emph{\textbf{Notation}}:
Boldface lower-case letters denote vectors.
For a random variable $X$, $\Pr\{\cdot\}$ denotes its probability, and $\mathbb{E}\{\cdot\}$ represents its expectation.
$\mathbf{I}_m$ is the $m \times m$ identity
matrix.
For a random vector variable $\mathbf{n}$,
$\mathbf{n}\sim\mathcal{CN}(0, \mathbf{\Omega})$ denotes a circular symmetric complex Gaussian variable with zero mean and covariance matrix $\mathbf{\Omega}$.
$\bigoplus$ represents XOR operation  and $Q(x)$ is the $Q$-function, given by, $Q(x)=\frac{1}{\sqrt{2\pi}}\int_x^{\infty}e^{-t^2/2}\,\mathrm{d}t$.

%% Introduction section
%% ==================================== %%

%% ==================================== %%
%% Model section
% \section{System Model}\label{sec:system.model}

\section{Compressed Relaying Scheme via  Huffman and Physical-Layer Network Coding}\label{sec:HPNC}
\begin{figure}[]%[e]%[!t]
\centering
%\graphicspath{{fig/}} %width=0.38
\includegraphics[width=0.55\textwidth]{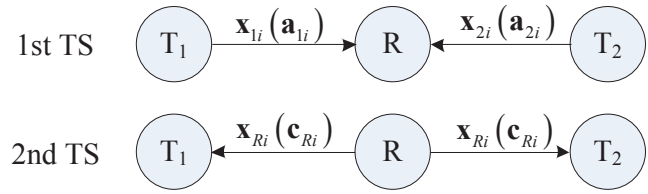}
\caption{The block diagram of the two-hop  TWRN-CS with the HPNC scheme.}
\label{fig:TWRN_DF_3ts_HPNC}
\end{figure}

A general two-hop TWRN-CS with three nodes is considered in this study, as shown in Fig.~\ref{fig:TWRN_DF_3ts_HPNC}, where two correlated source nodes, denoted by $T_1$ and $T_2$, want to exchange correlated messages with each other via a relay node, denoted by $R$.
All nodes are equipped with one antenna and operate in a half-duplex mode.
We assume that all the links between the sources and relay node are additive white Gaussian noise (AWGN) channels and the transmit powers of the three nodes are equal, denoted by $P_t$.

\subsection{Source Correlation Model}\label{subsec:SCM}
It is assumed that the information bits at both source nodes are divided equally into blocks and each block consists of $n$  bits.
Let $\mathbf{a}_{1i}=[a_{1i}(1), \cdots, a_{1i}(n)]$ and $\mathbf{a}_{2i}=[a_{2i}(1), \cdots, a_{2i}(n)]$ be the $i$th block of message  transmitted by $T_1$ and $T_2$,   where $\mathbf{a}_{1i}$, $\mathbf{a}_{2i} \in \{0,1\}^{1\times n}$.
The binary phase-shift keying (BPSK) modulation is considered for simplicity and the modulated symbols of $\mathbf{a}_{1i}$ and $\mathbf{a}_{2i}$ are denoted by
$\mathbf{x}_{1i}=[x_{1i}(1), \cdots, x_{1i}(n)]$ and $\mathbf{x}_{2i}=[x_{2i}(1), \cdots, x_{2i}(n)]$,   where $\mathbf{x}_{1i}$, $\mathbf{x}_{2i} \in \{+1,-1\}^{1\times n}$.
The \emph{correlation factor} between two source nodes is  defined as
$r=\mathbb{E}\{x_{1i}(j) x_{2i}(j)\}$, $\forall i$ and $1\leq j \leq n$.

\subsection{Compressed Relaying  via  Huffman and Physical-Layer Network Coding}\label{subsec:xx.yy}

\subsubsection{Encoding at the Relay}\label{subsec:xx.yy.zz}
In the proposed HPNC scheme, both source nodes transmit the BPSK modulated symbols of $\mathbf{a}_{1i}$ and $\mathbf{a}_{2i}$   to the relay, simultaneously, during the first time slot~(TS). Let $\mathbf{y}_{Ri}$   be the corresponding received signals at the relay and we have\cite{Yang2012P1499}
\begin{equation}
\mathbf{y}_{Ri} = \sqrt{P_t}\mathbf{x}_{1i} + \sqrt{P_t}\mathbf{x}_{2i} + \mathbf{n}_{Ri},
\end{equation}
where $\mathbf{n}_{Ri}$  is the noise vectors  following $\mathcal{CN}(0, N_0 \mathbf{I}_{n})$.
Let $\gamma =\frac{P_t}{N_0}$   denote the  signal-to-noise ratio (SNR).

The relay first performs  PNC  on the received signals. The resulting PNC-coded message is represented by
  \begin{equation}%\label{}
   \begin{split}
 \mathbf{a}_{Ri,\text{PNC}}=\mathbf{PNC}(\mathbf{y}_{Ri}),
   \end{split}
  \end{equation}
where $\mathbf{a}_{Ri,\text{PNC}}$ is the estimation of $\mathbf{a}_{1i}\bigoplus\mathbf{a}_{2i}$, and $\mathbf{PNC}(\cdot)$ is the PNC mapping function in\cite{Liew2013P4}.
Then the relay performs Huffman coding on the PNC-coded message $\mathbf{a}_{Ri,\text{PNC}}$.
Let $\mathbf{Huffman}: \mathcal{A}  \rightarrow \mathcal{C}$ and $\mathbf{invHuffman}: \mathcal{C} \rightarrow \mathcal{A}$ denote the standard Huffman encoding and decoding algorithm \cite{Huffman1952P1098}, where $\mathcal{A}=\{0,1\}^{1\times n}$ denotes the set of $\mathbf{a}_{Ri,\text{PNC}}$  and $\mathcal{C}$ denotes the set of the codewords of the corresponding Huffman code.
The compressed messages, denoted as $\mathbf{c}_{Ri}$, are given by
  \begin{equation}%\label{}
   \begin{split}
 \mathbf{c}_{Ri}
      &=\mathbf{Huffman}(\mathbf{a}_{Ri,\text{PNC}}).
   \end{split}
  \end{equation}
It should be mentioned that the block length of the compressed messages $\mathbf{c}_{Ri}$, denoted by $n_i$, is dependent on both the Huffman code design  and the PNC-coded message  $\mathbf{a}_{Ri,\text{PNC}}$.

During the second TS, $\mathbf{c}_{Ri}$ is modulated into $\mathbf{x}_{Ri}$  and then sent   to  both $T_1$ and $T_2$. The corresponding received signal vectors at  $T_1$ and $T_2$, denoted by $\mathbf{y}_{Ri,1}$ and $\mathbf{y}_{Ri,2}$,  can be written as
\begin{equation}
\mathbf{y}_{Ri,1} = \sqrt{P_t}\mathbf{x}_{Ri} + \mathbf{n}_{Ri,1}
\quad \text{and} \quad
\mathbf{y}_{Ri,2} = \sqrt{P_t}\mathbf{x}_{Ri} + \mathbf{n}_{Ri,2},
\end{equation}
where $\mathbf{n}_{Ri,1} \sim\mathcal{CN}(0, N_0 \mathbf{I}_{n_i})$ and  $\mathbf{n}_{Ri,2} \sim\mathcal{CN}(0, N_0 \mathbf{I}_{n_i})$ are the noise vectors experienced at $T_1$ and $T_2$, respectively.

\subsubsection{Decoding at the Destination Nodes}\label{subsubsec:xx.yy.zz}
In the following,  the decoding algorithm and the performance analysis for the signal flow from $T_1$ to $T_2$ are presented. Due to the symmetry, the discussions for the signal flow from $T_2$ to $T_1$ are  omitted   for brevity.
Upon receiving the compressed messages from the relay,  $T_2$ first calculates the hard estimation of $\mathbf{c}_{Ri}$, denoted by $\hat{\mathbf{c}}_{Ri,2}$.
Then $T_2$   can  use its own message, i.e., $\mathbf{a}_{2i}$, as side information to recover the desired message bits sent from $T_1$, denoted by $\hat{\mathbf{a}}_{1i,2}$.

Let us first consider an ideal case in which all nodes can decode the messages correctly.
In the ideal case, we have $\hat{\mathbf{c}}_{Ri,2}
=\mathbf{Huffman}(\mathbf{a}_{1i}\bigoplus\mathbf{a}_{2i})$.
$T_2$ can
perfectly construct $\mathbf{a}_{Ri,\text{PNC}}$  by decoding $\hat{\mathbf{c}}_{Ri,2}$, i.e., $\mathbf{a}_{Ri,\text{PNC}}=\mathbf{a}_{1i}\bigoplus\mathbf{a}_{2i}= \mathbf{invHuffman}(\hat{\mathbf{c}}_{Ri,2})$.
Finally, the desired message bits $\mathbf{a}_{1i}$ can be perfectly recovered at  $T_2$ as
$     \hat{\mathbf{a}}_{1i,2}
      = \mathbf{a}_{2i}\bigoplus (\mathbf{a}_{1i}\bigoplus\mathbf{a}_{2i})
      = \mathbf{a}_{2i}\bigoplus \mathbf{a}_{Ri,\text{PNC}}
      = \mathbf{a}_{2i}\bigoplus\mathbf{invHuffman}(\hat{\mathbf{c}}_{Ri,2})
$.
The decoding process can be similarly extended to the non-ideal case by taking into account decoding errors at the relay and destination.
Similar to the ideal case, $T_2$  first  constructs  the  estimated symbol vector of  $\mathbf{a}_{Ri,\text{PNC}}$, i.e., $\hat{\mathbf{a}}_{Ri,\text{PNC}}$, by decoding the  compressed message received from the relay node, i.e., $\hat{\mathbf{c}}_{Ri,2}$.
Then the estimated information bits sent from  $T_1$ are constructed as
\begin{equation}%\label{}
   \begin{split}
     \hat{\mathbf{a}}_{1i,2}
     &= \mathbf{a}_{2i}\bigoplus \hat{\mathbf{a}}_{Ri,\text{PNC}}
     = \mathbf{a}_{2i}\bigoplus\mathbf{invHuffman}(\hat{\mathbf{c}}_{Ri,2}).
   \end{split}
  \end{equation}

\subsection{Huffman Code Design}\label{subsec:xx}
In the following, the details on the design of  Huffman code for the proposed HPNC scheme are presented. Since the Huffman coding algorithm is uniquely determined  by the distribution of its input source symbols, we need to derive  the distribution of the PNC-coded message, i.e.,  $\mathbf{a}_{Ri,\text{PNC}}$.
In the HPNC scheme, we use the distribution of
$\mathbf{b}_i=\mathbf{a}_{1i}\bigoplus\mathbf{a}_{2i}$ as an approximated  distribution of
$\mathbf{a}_{Ri,\text{PNC}}$ to construct the Huffman codes.
It should be mentioned that the difference between the distribution of $\mathbf{b}_i$  and $\mathbf{a}_{Ri,\text{PNC}}$
is negligible especially at medium and high SNR values.
For simplicity, we define the \emph{equal factor} as $\rho = \Pr\{a_{1i}(j)=a_{2i}(j)\}= \Pr\{x_{1i}(j)=x_{2i}(j)\}$, $\forall i$ and $1 \leq j \leq n$. According to the source correlation model in Subsection \ref{subsec:SCM}, it is easy to derive  the relationship between the equal factor and the correlation factor as  $\rho=(r+1)/2$. The probability of $\mathbf{b}_i$ can be expressed as
\begin{equation}\label{eq:Pr_n_i}
\begin{split}
\Pr\{\mathbf{b}_i\}&=\rho^{m(\mathbf{b}_i)}(1-\rho)^{n-m(\mathbf{b}_i)}\\
&=2^{-n}(1+r)^{m(\mathbf{b}_i)}(1-r)^{n-m(\mathbf{b}_i)},
\end{split}
\end{equation}
where $m(\mathbf{b}_i)$ denotes the number of $0$ in $\mathbf{b}_i$, i.e., the number of $j$, $1\leq j\leq n$, that satisfies $a_{1i}(j)=a_{2i}(j)$.
Based on the derived approximated distribution, i.e.,  Eq. (\ref{eq:Pr_n_i}), we can perform Huffman encoding and decoding algorithm using the standard process in\cite{Huffman1952P1098}.
%% Model section
%% ==================================== %%

%% ==================================== %%
%% Analysis section

\section{Performance Analysis}\label{sec:analysis}

\subsection{Compression Rate}\label{subsec:compression_rate}

As indicated in Sec. \ref{subsec:xx}, the  Huffman code design is dependent on both the block length $n$ and the correlation factor $r$.
Let $\mathcal{C}(n,r)$ denote the set of the Huffman codeword, and  $l(\mathbf{b}_i)$ denote the codeword length of $\mathbf{b}_i$  after  Huffman encoding.
Let $\mathcal{LC}(n,r)$  be the range of $l(\mathbf{b}_i)$ and thus we have $l(\mathbf{b}_i)\in\mathcal{LC}(n,r),   \forall \mathbf{b}_i \in \{0,1\}^{1\times n}$. The codeword length distribution of the $i$th block messages, denoted by $\Pr\{n_i=k\}$, can be expressed as
$\Pr\{n_i=k\}=\sum_{\mathbf{b}_i:l(\mathbf{b}_i)=k}\Pr\{\mathbf{b}_i\} =\sum_{\mathbf{b}_i:l(\mathbf{b}_i)=k}2^{-n}(1+r)^{m(\mathbf{b}_i)}(1-r)^{n-m(\mathbf{b}_i)},
$
where $k \in \mathcal{LC}(n,r)$.
Thus, the average codeword length is given by
$\bar{n}_i  = \mathbb{E}\{n_i\} = \sum_{k \in \mathcal{LC}(n,r)}k\cdot\Pr\{n_i=k\}$.

Since the raw source messages are transmitted without compression at both sources during the first TS, the overall compression rate of the HPNC scheme is given by
\begin{equation}\label{eq:bar_n_i}
\begin{split}
C_\text{HPNC}=\frac{n+\bar{n}_i}{2n}=\frac{1}{2}+\frac{\bar{n}_i}{2n}.
\end{split}
\end{equation}
Note that the theoretical maximum compression rate at the relay node, i.e., entropy rate, can be readily derived as
$C_\text{Theo,R}=H[\mathbf{a}_{1i}|\mathbf{a}_{2i}]/n =H[\frac{1+r}{2}]= -\frac{1+r}{2}\log(\frac{1+r}{2})-(\frac{1-r}{2})\log(\frac{1-r}{2}) $\cite{Cover2006P}.
According to the property of  Huffman coding, we have   $nH[\frac{1+r}{2}] \leq \bar{n}_i  <nH[\frac{1+r}{2}] +1$  \cite{Cover2006P}. Thus  we can  obtain $C_\text{Theo} \leq C_\text{HPNC}< C_\text{Theo} + 1/2n$, where $C_\text{Theo}=1/2+H[\frac{1+r}{2}]/2$. It implies that when the number of symbols within one block is large, the compression rate of the HPNC scheme $C_\text{HPNC}$ approaches the theoretical maximum compression rate $C_\text{Theo}$ and the gap is no more than $1/2n$.

\subsection{Block Error Rate}\label{subsec:xx.yy}

The decision error at the first TS can be calculated as  
\begin{equation}
\begin{split}
\label{eq:ser_PNC}
P_{\text{PNC}}(\gamma,\rho) &= (1-\rho) \hspace{-1.5mm} \int\limits_{\tau_1}^{\infty} \hspace{-1.5mm} \sqrt{\frac{1}{\pi N_0}}e^{-\frac{y^2}{N_0}}  d y + (1-\rho) \hspace{-1.5mm} \int\limits_{-\infty}^{-\tau_2} \hspace{-1.5mm} \sqrt{\frac{1}{\pi N_0}}e^{-\frac{y^2}{N_0}} d y \\
& \qquad \qquad
+ \frac{\rho}{2} \hspace{-1.5mm} \int\limits_{-\tau_2}^{\tau_1} \hspace{-1.5mm} \sqrt{\frac{1}{\pi N_0}}e^{-\frac{(y-2\sqrt{P_t})^2}{N_0}}  d y + \frac{\rho}{2} \hspace{-1.5mm} \int\limits_{-\tau_2}^{\tau_1} \hspace{-1.5mm} \sqrt{\frac{1}{\pi N_0}}e^{-\frac{(y+2\sqrt{P_t})^2}{N_0}}   d y, 
\end{split}
\end{equation} 
where   $\tau_1$ and  $-\tau_2$  are the decision threshold values. 
According to the maximum posterior probability criterion and the symmetrical property, we can obtain the optimal decision threshold as
$\tau_1 = \tau_2 = \tau =\sqrt{P_t}+ \frac{\sqrt{N_0}}{4\sqrt{\gamma}}\ln \frac{1-\rho}{\rho} + \frac{\sqrt{N_0}}{4\sqrt{\gamma}}\ln\left(1+\sqrt{1-(\frac{\rho}{1-\rho})^2 e^{-8\gamma}}\right)$ when $\frac{1-\rho}{\rho} e^{4\gamma}>1$, otherwise we have $\tau_1 = \tau_2 = \tau =0$.
The optimal decision rule at the relay is described as follows:
when the received signal is less than $-\tau$, we declare $x_{1i}(j)+x_{2i}(j)$ to be $-2$; when the received signal is larger than $\tau$, we declare $x_{1i}(j)+x_{2i}(j)$ to be $2$; otherwise, it is set to be $0$.
According to Eq.~(\ref{eq:ser_PNC}), the error probability of the PNC mapping with  the optimal decision threshold over single symbol can be expressed as
\begin{align}
&P_{\text{PNC}}(\gamma, \rho) = 2(1-\rho)Q\left(\bar{\tau}\right) + \rho Q\left(2\sqrt{2\gamma}-\bar{\tau}\right)  - \rho Q\left(2\sqrt{2\gamma}+\bar{\tau}\right),
\end{align}
where
$\bar{\tau}= \sqrt{\frac{2}{N_0}}\tau$.
Therefore, the error probability  of  the PNC mapping during the first TS can be expressed as
\begin{equation}\label{eq:T1_R_BLER}
\begin{split}
  P_{R}(\gamma, \rho)   = 1-\left(1-P_{\text{PNC}}(\gamma, \rho)\right)^{n}.\\
\end{split}
\end{equation}

Let $P_{R2}(\gamma|k)$ denote the error probability of the transmission of  $\mathbf{c}_{Ri}$  from $R$ to $T_2$ conditioned on  the length of the codeword $\mathbf{c}_{Ri}$, denoted by $k$, then we have
$P_{R2}(\gamma|k)  = 1-(1-Q(\sqrt{2\gamma}))^{k}$.
The average BLER of the transmission from $R$ to $T_2$ is given by
\begin{equation}\label{eq:R_T1_BLER_comp}
\begin{split}
  P_{R2}(\gamma)  & = \sum_{k \in \mathcal{LC}(n,r)}\Pr\{n_i=k\} \cdot P_{R2}(\gamma|k).  \\
\end{split}
\end{equation}

Let $P_{\text{HPNC},12}$ denote the BLER of the transmission from $T_1$ to $T_2$ in the HPNC scheme. A block error, i.e., $\hat{\mathbf{a}}_{1i,2} \neq \mathbf{a}_{1i}$, only occurs if the received compressed message is erroneous, i.e., $\hat{\mathbf{c}}_{Ri,2} \neq \mathbf{Huffman}(\mathbf{a}_{1i}\bigoplus\mathbf{a}_{2i})$.
Thus, $P_{\text{HPNC},12}$ can be calculated as
\begin{equation}\label{eq:T1_T2_BLER}
\begin{split}
   & P_{\text{HPNC},12}(\gamma))
   = 1-\left(1-P_{R}(\gamma, \rho))\right) \left(1-P_{R2}(\gamma)\right). \\
\end{split}
\end{equation}
By applying the approximation $1-(1-x)^{N}\approx N x$ when $x$ is small.
At high SNR regime, Eq. (\ref{eq:T1_T2_BLER})  can be approximated as
\begin{equation}\label{eq:T1_T2_BLER_approx_lowSNR}
\begin{split}
  P_{\text{HPNC},12}(\gamma)
  \approx   nP_{\text{PNC}}(\gamma, \rho) + \bar{n}_iQ(\sqrt{2\gamma}).
\end{split}
\end{equation}
At high SNR regime, the threshold $\bar{\tau}$ approaches to $\sqrt{2\gamma}$, and $P_{\text{HPNC}}(\gamma, \rho) \approx (2-\rho)Q(\sqrt{2\gamma})$.
The BLER can be calculated approximately as
\begin{equation}\label{eq:T1_T2_BLER_approx} %{eq:hpnc_apr}
P_{\text{HPNC},12}(\gamma) \approx (2n-\rho n+\bar{n}_i) Q(\sqrt{2\gamma}).
\end{equation}

Similarly, the exact and asymptotic  BLER expressions  of the conventional non-compressed relaying scheme are derived as
\begin{equation}\label{eq:T1_T2_BLER_exact_trad}
\begin{split}
    P_{\text{Conv},12}(\gamma)
     =& 1-\left(1-P_{R}(\gamma, 0.5)\right)  \left(1-P_{R2}(\gamma|n)\right),
\end{split}
\end{equation}
and
\begin{equation}\label{eq:T1_T2_BLER_approx_trad}
\begin{split}
   P_{\text{Conv},12}(\gamma)
     \approx & \frac{5n}{2} Q(\sqrt{2\gamma}),
\end{split}
\end{equation}
respectively.
From Eqs. (\ref{eq:T1_T2_BLER_approx}) and (\ref{eq:T1_T2_BLER_approx_trad}),
the gain of the proposed scheme over the conventional scheme in terms of  the ratio of BLER at high SNR regime yields
\begin{equation} \label{eq:BLER_gain}
\begin{split}
G_{\text{BLER}}=\lim_{\gamma \rightarrow\infty}\frac{P_{\text{Conv},12}(\gamma)}{P_{\text{HPNC},12}(\gamma)} =  \frac{5}{4C_\text{HPNC}+2-2\rho}.
\end{split}
\end{equation}
Eq. (\ref{eq:BLER_gain})  indicates  that the  proposed HPNC scheme can achieve considerable BLER performance improvement  with respect to the conventional scheme when the compression rate $C_\text{HPNC}$ is small.

%% Analysis section
%% ==================================== %%

%% ==================================== %%
%% Simulations section

\section{Simulation Results}\label{sec:simulations}
In this section,   analytical and simulated results for the HPNC scheme in TWRN-CS are presented.
Simulation results are conducted for a  BPSK modulation  and a block size of $n=6$  over AWGN channels.  Five cases with different correlation factors between $T_1$ and $T_2$ are considered: 1) $r=0.9$; 2) $r=0.8$; 3) $r=0.7$; 4) $r=0.6$;  and 5) $r=0.4$.

\begin{figure}[]%[e]%[!t]
\centering
%\graphicspath{{fig/}}
\includegraphics[width=0.65\textwidth]{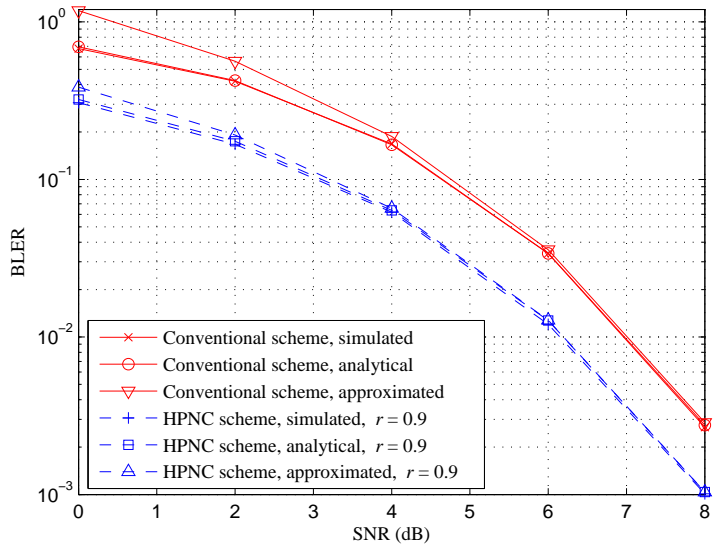}
\caption{BLER comparison between the proposed HPNC scheme  and the conventional non-compressed relaying scheme with $n=6$.}
\label{fig:corr_Huf_BLER}
\end{figure}

\begin{figure}[]%[e]%[!t]
\centering
%\graphicspath{{fig/}}
\includegraphics[width=0.65\textwidth]{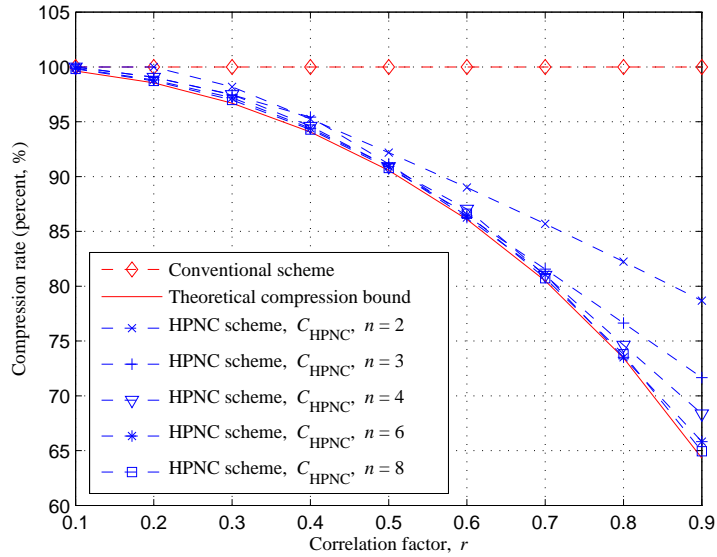}
\caption{Compression rates comparison between the proposed HPNC scheme and the conventional non-compressed relaying scheme.}
%\caption{Compression rates at the relay.}
\label{fig:corr_Huf_rate}
\end{figure}

In Fig. \ref{fig:corr_Huf_BLER},  the  analytical and simulated BLER performance of the HPNC scheme and the conventional non-compressed relaying scheme are presented.
It shows that the  analytical theoretical curves obtained from Eqs. (\ref{eq:T1_T2_BLER}) and (\ref{eq:T1_T2_BLER_exact_trad})  match  perfectly with the simulated results, and the asymptotic results computed in Eqs. (\ref{eq:T1_T2_BLER_approx_lowSNR}) and  (\ref{eq:T1_T2_BLER_approx_trad}) converge to the simulated results at high SNRs.  This validates the accuracy of the BLER analysis.
From the comparisons in Fig.~\ref{fig:corr_Huf_BLER}, we can observe that the proposed scheme is superior to the conventional scheme.

Fig. \ref{fig:corr_Huf_rate} compares the analytical compression rate of the HPNC scheme   $C_\text{HPNC}$  in terms of the different block size $n$. The corresponding theoretical maximum compression rate $C_\text{Theo}$ is also given in this figure. It shows that
the compression rate of the proposed scheme approaches
the theoretical maximum compression rate quickly when the block size of the information bits, i.e., $n$,  increases, which validates our analysis of compression rate given in Sec. \ref{subsec:compression_rate}.

\begin{figure}[]%[e]%[!t]
\centering
%\graphicspath{{fig/}}
\includegraphics[width=0.65\textwidth]{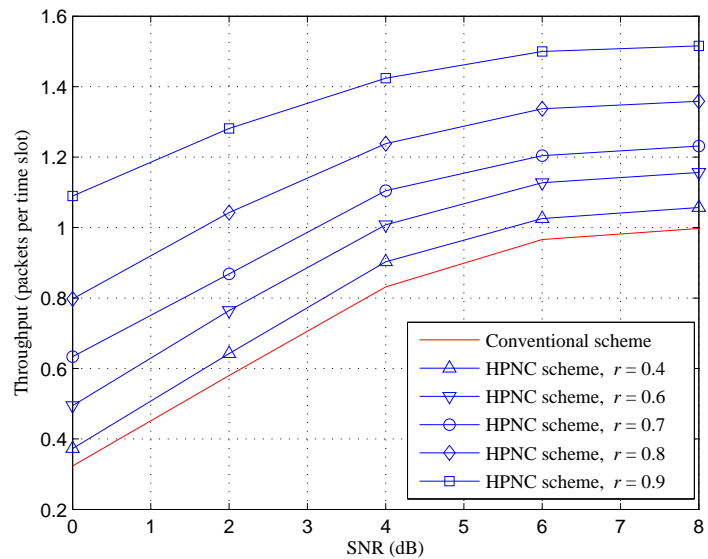}
\caption{Throughput comparison between the proposed HPNC scheme  and the conventional non-compressed relaying scheme with $n=6$.}
\label{fig:corr_Huf_throughput}
\end{figure}

In Fig. \ref{fig:corr_Huf_throughput}, we compare the simulated network throughput for the proposed scheme and the conventional non-compressed relaying scheme.
Define the throughput as the number of  message blocks which are decoded correctly at source node $T_1$ and $T_2$ per TS.
It can be observed  that the  proposed HPNC scheme  achieves considerable  throughput improvement  with respective to the conventional scheme.  It is also shown that the throughput increases significantly  as the correlation factor increases in the HPNC scheme. These gains come from the compressed  process at the relay  in TWRN-CS.

%% Simulations section
%% ==================================== %%

%% ==================================== %%
%% Conclusion section
\section{Conclusion }\label{sec:conclusion}
%The conclusion goes here.

In this letter, we has proposed a compressed relaying framework with jointly designing  Huffman  coding and PNC  for TWRN-CS.
Compression rate has been analyzed and analytical BLER  expression  has been derived   for the HPNC  scheme.
It has been  shown that the HPNC scheme  achieves  significant improvements   compared with the conventional non-compressed relaying scheme. The gain is  contributed from the compressed process at the relay.
The idea developed here for   AWGN channels can be easily extended to  fading channels; however, the theoretical analysis of the performance metrics in fading channels is non-trivial. This subject will be reported on by the authors in a forthcoming paper.

%% Conclusion section
%% ==================================== %%

%% ==================================== %%
%% reference section

\bibliographystyle{IEEEtran}
%\bibliography{IEEEabrv,Huo}

%% reference section
%% ==================================== %%

%% ==================================== %%
%% that's all folks

\end{document}